\documentclass{emulateapj}
\newcommand{\be}{\begin{equation}}
\newcommand{\etal}{{\em et al.}}
\newcommand{\ltorder}{\hbox{ \rlap{\raise 0.425ex\hbox{$<$}}\lower
0.65ex\hbox{$\sim$} }}
\newcommand{\gtorder}{\hbox{ \rlap{\raise 0.425ex\hbox{$>$}}\lower
0.65ex\hbox{$\sim$} }}

\shorttitle{Effects of primordial mass segregation on clusters evolution}
\shortauthors{E.Vesperini, S.L.W. McMillan, S. Portegies Zwart}
\submitted{Accepted for publication in The Astrophysical Journal}

\begin{document}

\title{Effects of Primordial Mass Segregation on the Dynamical Evolution of Star Clusters}
\author{Enrico Vesperini$^1$, Stephen L.W. McMillan$^1$, Simon Portegies Zwart $^{2,3}$ } 
\affil{$^1$
  Department of Physics, Drexel University, Philadelphia, PA 19104;
  vesperin@physics.drexel.edu,steve@physics.drexel.edu\\
$^2$ Astronomical Institute ``Anton Pannekoek'' and Section
  Computational Science, University of Amsterdam, Kruislaan 403,
  Amsterdam, The Netherlands.\\
$^3$ Sterrewacht Leiden, P.O. Box 9513, 2300 RA Leiden, The Netherlands; S.F.PortegiesZwart@uva.nl}

\begin{abstract}
In this paper we use $N$-body simulations to study the effects
of primordial mass  segregation on the early and long-term  evolution
of star clusters. Our
simulations show that in segregated clusters  early  mass loss due to
stellar evolution triggers a stronger expansion than for unsegregated
clusters. Tidally limited, strongly segregated clusters may dissolve rapidly as
a consequence of this early expansion, while segregated clusters
initially underfilling their 
Roche lobe can survive the early expansion and have a lifetime similar
to that of unsegregated clusters. Long-lived initially segregated clusters tend
to have looser structure  and reach core collapse later in their
evolution than initially unsegregated clusters.
We have also compared the effects of dynamical evolution on the global
stellar mass function (MF) of low-mass main sequence stars.  In all
cases the MF flattens as the cluster loses stars.  The amount of MF
flattening induced by a given amount of mass loss in a rapidly
dissolving initially segregated cluster is less than for an
unsegregated cluster.  The evolution of the MF of a long-lived
segregated cluster, on the other hand, is very similar to that of an
initially unsegregated cluster.

\end{abstract}
\section{Introduction}
\label{sec:intro}
Mass segregation, the tendency of more massive stars to preferentially
populate the inner parts of a star cluster, is one of 
the consequences of two-body relaxation and of the evolution toward
energy equipartition in stellar systems.
The characteristic timescale for this process, $T_{ms}$, is, for a
population of massive stars with mass $m_h$, $T_{ms}
\sim (\langle m \rangle/m_h) t_{relax}$ where $t_{relax}$ is the
cluster relaxation time and $\langle m \rangle$ the mean mass
of the cluster stars.

However, a number of young clusters with ages
significantly smaller than the time needed to produce the observed mass
segregation by standard two-body relaxation show a significant degree
of mass segregation (e.g.\ Hillenbrand 1997, Hillenbrand \& Hartmann
1998, Fischer \etal\ 1998, de Grijs \etal\ 2002, Sirianni \etal\
2002, Gouliermis \etal\ 2004, Stolte \etal\ 2006, Sabbi \etal\ 2008).

The results of a number of theoretical studies (e.g. Klessen 2001, \
Bonnell \etal\ 2001,  
Bonnell \& Bate 2006 but see also Krumholz \etal\ 2005, Krumholz \&
Bonnell 2007) showing that massive star would preferentially form
in the center of star-forming regions suggest that the observed
segregation in young clusters would be primordial and imprinted in a
cluster by  the star formation processes.
Possible dynamical routes leading to early mass segregation in young
star clusters have  been studied by McMillan, Vesperini \&
Portegies Zwart (2007). 

In this paper we present the results of a survey of $N$-body
simulations aimed at exploring the implications of primordial mass
segregation for the dynamical evolution of star clusters. We have
explored the evolution of clusters located at different galactocentric
distances and with different degrees of initial mass segregation and
compared the evolution of segregated clusters with that of clusters
with the same initial density profile but no initial mass
segregation. Our models focus on the effect of primordial mass
segregation on a purely stellar system; primordial gas and the effects
of its expulsion (see e.g. Baumgardt \& Kroupa 2007) are not included
here.

In Sect. \ref{sec:analytical} we present a preliminary analytical
calculation aimed at  exploring the differences in the response of
clusters with different initial degrees of mass segregation to  the
early  mass loss due to stellar evolution. This is, in most
cases,  the first process  affecting the cluster
structure well before the effects of two-body relaxation become
important.  We continue, in Sect.\ref{sec:nbody},  with the presentation of the
results of our $N$-body simulations, focussing our attention on clusters
lifetime, structural evolution and the evolution of the stellar
mass function.   We discuss and summarize our results in
Sect.\ref{sec:disc_and_concl}.   

\section{Early Evolution of Segregated Clusters: analytical estimates}
\label{sec:analytical}
A number of studies (see e.g. Chernoff \& Shapiro 1987, Chernoff \&
Weinberg 1990, Fukushige \& Heggie 1995, Takahashi \& Portegies Zwart
2000) have shown that early mass loss due to stellar evolution can
have a significant impact on the evolution of clusters: this early
mass loss causes a cluster to expand and, for a low-concentration
cluster, leads to the cluster's complete and rapid dissolution.  For
initially mass-segregated clusters, the mass lost due to the evolution
of massive stars is removed preferentially from the cluster's inner
regions, and the early expansion of the cluster is stronger and
potentially more destructive than when the same amount of mass is lost
in a non-segregated cluster.

Fig.~1 shows the results of a simple semi-analytical calculation
illustrating the augmented destructive effect of stellar mass loss in
a mass-segregated cluster.  In this figure we plot the virial ratio,
$T/V$, of an isolated cluster with an initial Plummer density profile 
$$
	\rho ={3M\over 4\pi a^3}\left(1+{r^2\over a^2}\right)^{-5/2}
$$
(see e.g. Heggie \& Hut 2003), after the impulsive loss of a fraction
$\Delta M$ of the total mass.  
\begin{figure}[h]
\begin{center}
\plotone{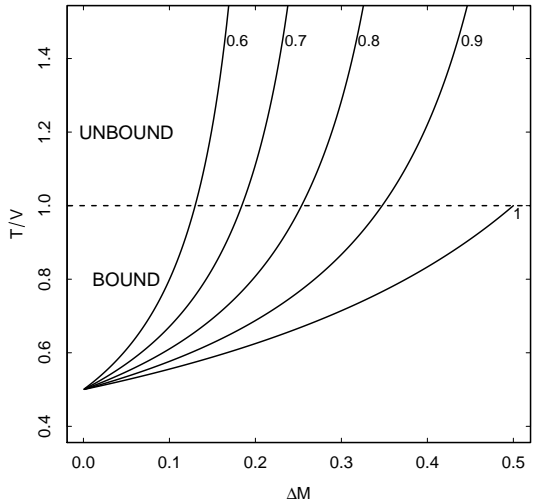} 
 \caption{Virial ratio of a cluster with a Plummer density profile
  after the rapid loss of a $\Delta M$ fraction of its initial
  mass, versus $\Delta M$. The density profile of the mass lost is
  assumed to be that of a Plummer model, but with a scale radius
  $a_{ML}$ equal to or less than the scale radius of the whole
  cluster, $a_{cluster}$. The number beside to each curve indicates
  the corresponding value of $ a_{ML}/a_{cluster}$}
   \label{fig1}
\end{center}
\end{figure}

In order to mimic the preferential mass loss from the inner regions of
a mass-segregated cluster, we have simply assumed that the density
profile of the mass lost also follows a Plummer model, but with a
scale radius ($a_{ML}$) smaller than the scale radius of the cluster
($a_{cluster}$).  The new potential ($V$) and kinetic ($T$) energies
for the density profile thus obtained are calculated.  The curves in
Fig.~1 show the resulting ratio $T/V$ as a function of $\Delta M$ for
different values of $a_{ML}/a_{cluster}$.

For $a_{ML}=a_{cluster}$ one recovers the well-known result that a
system becomes unbound for $\Delta M >0.5$ (Hills 1980; see Boily \&
Kroupa 2003 for a more detailed calculation showing that larger values
of the mass loss are actually necessary to unbind a cluster).
However, for a cluster preferentially losing mass from the inner
regions, $a_{ML}<a_{cluster}$, we see that a significantly smaller
amount of mass loss, innocuous for an unsegregated cluster, can lead
to the dissolution of a mass-segregated cluster.

This simple semi-analytical calculation underscores the potentially
crucial implication of initial mass segregation for the evolution of
star clusters; the results of the $N$-body simulations presented in the
rest of this paper illustrate these consequences in more detail.

\section{Dynamical evolution of Segregated Clusters: $N$-body simulations}
\label{sec:nbody}
\subsection{Method and Initial Conditions}
\label{sec:init}
The results presented in this section are based on a survey of direct $N$-body
simulations carried out using the {\tt starlab} package (Portegies
Zwart \etal\ 2001; {\tt http://www.manybody.org}) and accelerated by a GRAPE-6 
special-purpose board (Makino et al. 2003). 
\begin{figure}[h]
\begin{center}
\plotone{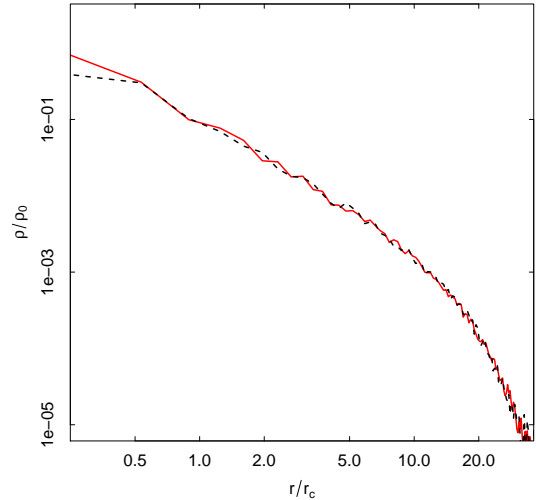} 
 \caption{Initial radial density profile (the radius is normalized to
 the core radius, $r_c$) for the {\it S} (solid red line)
 and the {\it U}  (dashed black line) clusters.}
\label{fig:densprof}
\end{center}
\end{figure}
\begin{figure}[h]
\begin{center}
 \plotone{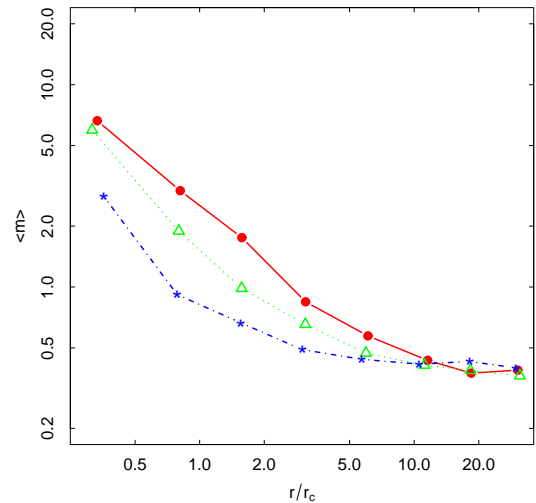} 
 \caption{Initial radial profile of the stellar mean mass for the
 initially segregated systems  investigated in our simulations: 
 {\it S} runs (solid red line/filled dots), {\it M} run (dotted green
 line/open triangles),  {\it L} run (blue dot-dashed line/asterisk
 dots).  
}
\label{fig:massprofile}
\end{center}
\end{figure}

 We have followed the evolution of clusters located at different
 galactocentric distances, $R_g=1, 4, 18, 40$ kpc, in a
 Galactic tidal 
field modeled as a Keplerian potential determined by a point mass
$M_g$ equal to the total mass of the Galaxy inside the distance $R_g$. 
We have studied both tidally truncated clusters and clusters
 underfilling their Roche lobe with a ratio $R/R_t$ of the total radius, $R$ to the
 tidal radius, $R_t$ equal to 0.75 and 0.5.
A Kroupa  (Kroupa, Tout \& Gilmore 1993) stellar initial mass function
 with star masses 
between 0.1 and 100 $m_{\odot}$ was adopted and star masses were
generated by the analytical function in Eq. 14 of Kroupa et al. (1993).

For segregated systems,  mass segregation
was  set up by first letting the cluster evolve without including the
effects of stellar evolution until a given degree of segregation was
reached due to normal two-body relaxation. 
We emphasize that this is
just the procedure we have used to generate a self-consistent
initially segregated cluster and it is not meant to model any stage of
cluster evolution. Alternative procedures  to produce initially
segregated clusters have recently been suggested (see e.g. Subr \etal\
2008, Baumgardt \etal\ 2008).

Fig.~2 and Fig.~3 show, respectively, the initial density profile  and
the initial radial profile of the star mean mass for the systems we
have studied;  in particular, the solid
line in Fig.~\ref{fig:massprofile} shows the initial  
degree of mass segregation of the standard models on which we focus 
in this paper.
The initial number of particles for the standard segregated ($S$)
and unsegregated ($U$) runs is $N=25$K. Initial conditions for
moderate and low initial mass segregation (the {\it M-Rg18} and {\it
  L-Rg18} runs, respectively)  are obtained 
by extracting snapshots at earlier times of  the same preliminary 
simulation mentioned above when the system has a lower degree of
segregation and has lost a smaller number of particles; the initial
number of particles for the {\it M-Rg18} and the {\it L-Rg18} runs are,
respectively, $N=33$K and $N=38$K. 

In order to explore the dependence on $N$ of the evolution of clusters most
affected by initial mass segregation,  simulations with the same degree
of initial segregation as {\it S-Rg18},
 {\it M-Rg18}  and {\it L-Rg18} have been repeated with a larger
 number of particles ($N=70$K, $N=103$K and $N=120$K respectively). 
\begin{table}[h]
\begin{center}
\begin{tabular}{|l|c|c|c|}
\hline
Id. & $R_g (kpc)$ & $R/R_t$ & Initial mass segregation\\
\hline
S-Rg1 & 1 & 1 &  strong\\
S-Rg4 & 4 & 1 &  strong\\
S-Rg18 & 18 & 1 &  strong\\
S-Rg18R075 & 18 & 0.75 &  strong\\
S-Rg18R05 & 18 & 0.5 &  strong\\
S-Rg40 & 40 & 1 &  strong\\
\hline
U-Rg1 & 1 & 1 &  no segregation\\
U-Rg4 & 4 & 1 &  no segregation\\
U-Rg18 & 18 & 1 &  no segregation\\
U-Rg18R075 & 18 & 0.75 &  no segregation\\
U-Rg18R05 & 18 & 0.5 &  no segregation\\
U-Rg40 & 40 & 1 &  no segregation\\
\hline
M-Rg18 & 18 & 1 &  moderate\\
L-Rg18 & 18 & 1 &  low\\
\hline
\hline
\end{tabular}
\end{center}
\caption{Initial conditions of all simulations.  The columns list (1)
Id (Standard Segregated Run-{\it S}, Standard Unsegregated
Run-{\it U},Moderate Segregation Run-{\it M}, Low Segregation Run-{\it
  L}),(2)
galactocentric distance, $R_g$ in kpc, (3) ratio of the initial cluster
radius to the tidal radius $R/R_t$, (4) degree of initial mass
segregation} 
\end{table}

We summarize in Table 1 all the initial conditions considered,
along with the identification used throughout the paper to refer to
each simulation.  

\subsection{Results: cluster evolution and lifetime}
\label{sec:lifetime}
The analytical calculation presented in Sect.\ref{sec:analytical}
showed that early impulsive mass loss associated with  
stellar evolution can have a stronger impact on initially segregated
cluster than unsegregated clusters.

For tidally truncated clusters with fixed initial mass,
evolving in a host galaxy with constant circular velocity, 
the mean cluster density $\rho_{tid}
\propto M/R_{t}^3 \propto M_g/R_g^3$ decreases with the galactocentric
distance $R_g$ as $\rho_{tid} \propto 1/R_g^2$.   The  
cluster dynamical time, defined using the mean density within the
tidal radius, $t_{tid}\propto 1/\sqrt{\rho_{tid}}$ thus increases linearly
with $R_g$
(note that this dynamical time is proportional to the circular orbital
period of the cluster around the Galaxy at distance $R_g$).
This scaling implies
that, for tidally truncated clusters, the amount of impulsive mass
loss due to stellar evolution increases with $R_g$.

\begin{figure}[h]
\begin{center}
\epsscale{0.95}
 \plotone{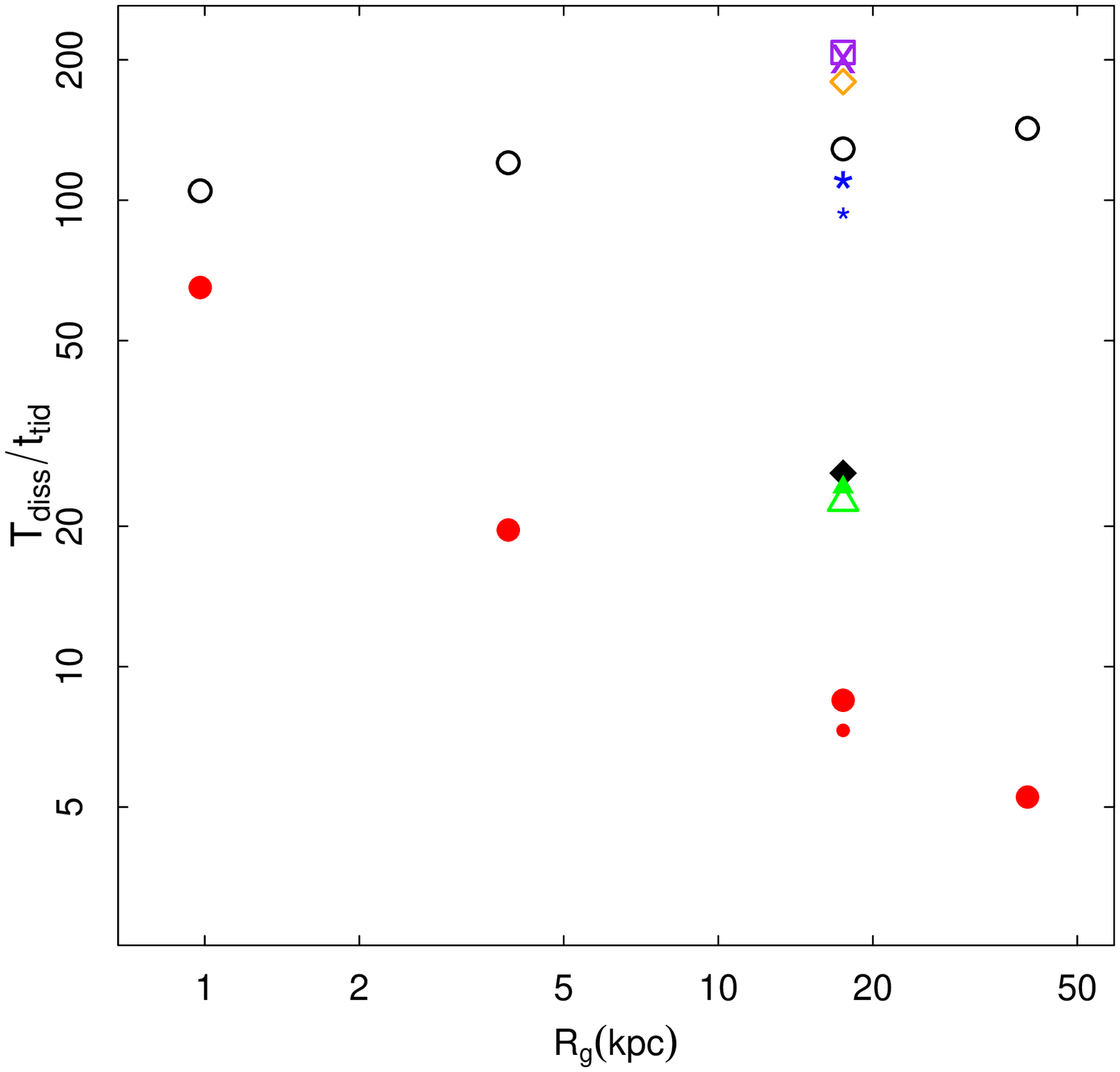}\\
 \plotone{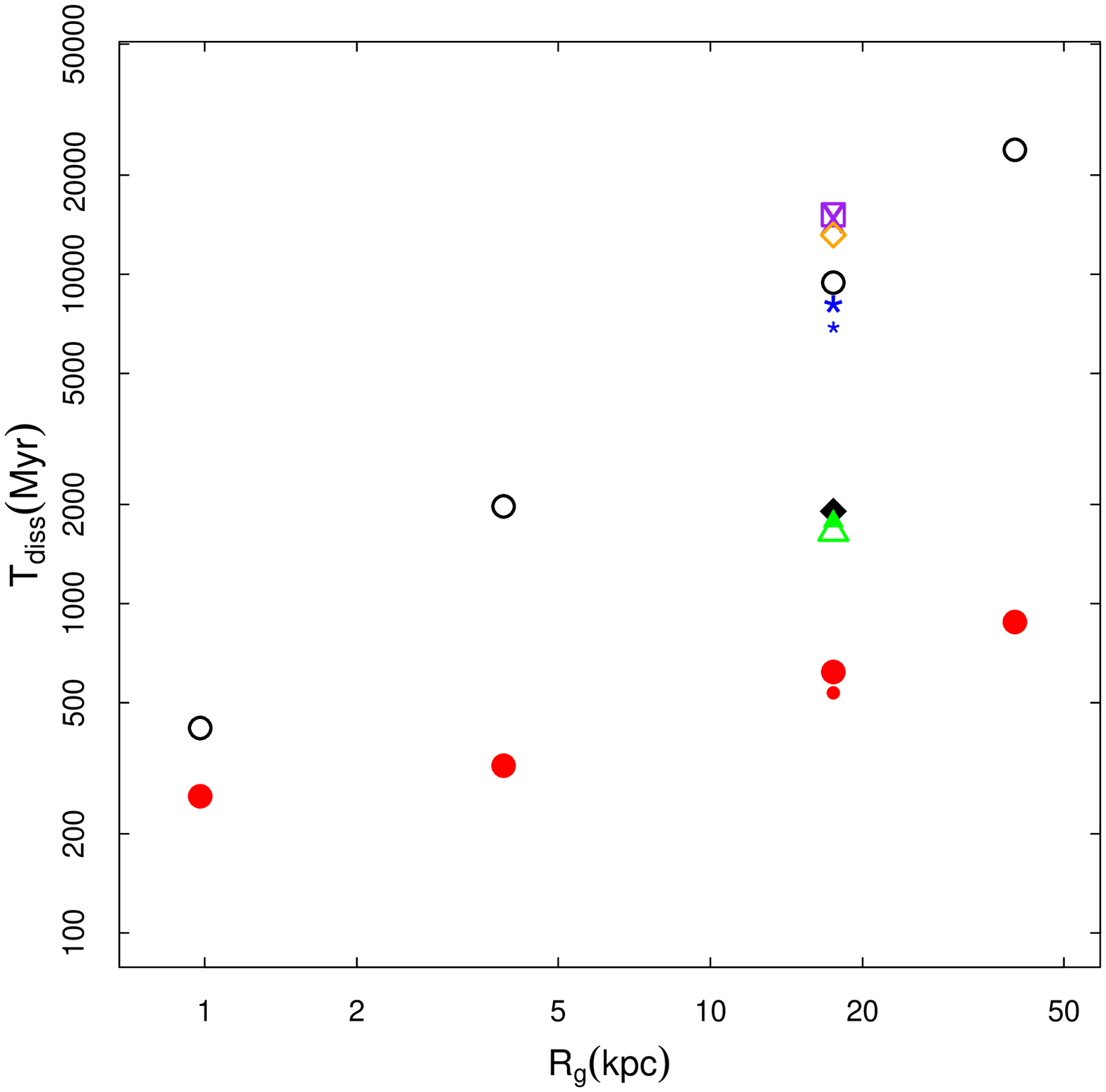}
 \caption{({\it Upper panel}) Dissolution time, $T_{diss}$, (in units
   of cluster dynamical time, $t_{tid}$) versus galactocentric distance, $R_g$,
   for  the  segregated (red filled
   dots-{\it S} runs) and unsegregated clusters (black open dots-{\it U}
   runs). The green open triangle and the blue asterisk-dot show, respectively, the
   dissolution time for the 
   moderate-segregation ({\it M-Rg18}) and low-segregation ({\it L-Rg18})
   runs. The black filled diamond-dot and the open orange diamond-dot
   show, respectively, the dissolution times
   for {\it S-Rg18R075} and {\it U-Rg18R075}; the purple cross
   and the purple square-dot (almost overlapping) show, respectively, the dissolution times 
   for {\it S-Rg18R05} and {\it U-Rg18R05}  (for a consistent
   comparison of all the models at $R_g=18$ kpc, the dynamical time
   used to normalize the dissolution times of {\it S-Rg18R075},
   {\it U-Rg18R075}, {\it S-Rg18R05} and {\it U-Rg18R05} is the 
   same used for the other models at $R_g=18$ kpc and equal to the
   dynamical time of the tidally truncated systems). The small red
   filled dot, the small green filled triangle (overlapping with the
   large triangle)  and the small blue asterisk-dot
   show the dissolution
   times for {\it S-Rg18}, {\it M-Rg18} and {\it L-Rg18}  repeated with a
   larger number of particles ($N=70$K, $N=103$K and $N=120$K
   respectively). ({\it Lower  panel}) Dissolution time (in Myr)
   versus galactocentric distance  (symbols as in the upper panel).} 
\label{fig:tdissol} 
\end{center}
\epsscale{1}
\end{figure}

Fig.~\ref{fig:tdissol} shows the
dependence  of the cluster dissolution time, $T_{diss}$, (defined as
the time when 1 percent of the initial mass is left in the cluster)
on the cluster galactocentric distance for
mass-segregated and  unsegregated clusters. At a given $R_g$, the amount of
impulsive mass loss is the same for both segregated and unsegregated
clusters but, as anticipated by the analytical calculations presented
earlier,  the destructive effect of this mass loss is
strongly augmented by mass segregation.

Fig.~\ref{fig:tdissol} shows that all of the unsegregated clusters survive 
the early mass loss due to stellar evolution and eventually
dissolve as a result of the evaporation of stars due to two-body
relaxation. On the other hand, early impulsive mass loss leads to the
dissolution within just a few dynamical times of all of the initially
mass-segregated clusters. Only the dissolution time of the cluster
closest to the 
Galactic center (for which the dynamical time is short  and the
impulsive mass loss is negligible) is similar to that of the corresponding
unsegregated cluster.  

\begin{figure}[h]
\begin{center}
 \plotone{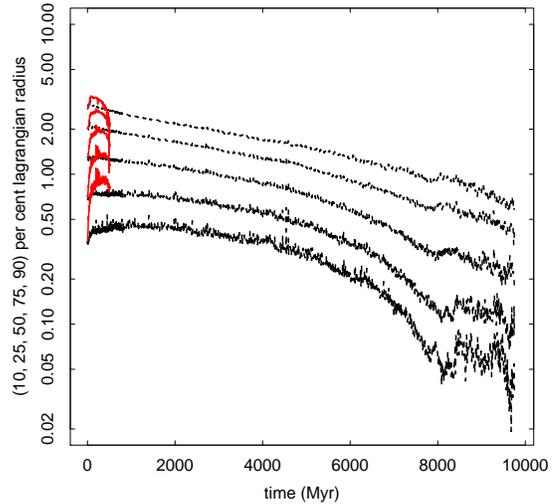} 
 \caption{Time evolution of the 10, 25, 50, 75, 90 percent Lagrangian
 radii for {\it S-Rg18}  (red solid line) and {\it U-Rg18} (black dashed
 line).}
   \label{fig:lagrRg18}
\end{center}
\end{figure}

As already noted in a number of previous studies (see
e.g. Vesperini \& Heggie 1997, Portegies Zwart et al. 1998, Baumgardt \& Makino 2003), the
dissolution time of tidally truncated unsegregated clusters increases
linearly with $R_g$.  This dependence is evident from the lower panel
of Fig.\ref{fig:tdissol}; in the upper panel it is
incorporated in the dynamical time, $t_{tid}$, used  to normalize the
$T_{diss}$  (as we pointed out above in this section,
$t_{tid} \propto R_g$).  
The dissolution of the unsegregated clusters is driven principally by two-body
relaxation, and for clusters more massive than those we have studied
in our simulations, the ratio $T_{diss}/t_{tid}$ increases with the
initial number of stars $N$ as $\sim (N/\log N)^x$ with $x\sim
0.75-0.8$ (see Baumgardt \& Makino 2003).  

For initially mass-segregated systems the dependence of $T_{diss}$ on $R_g$
is determined by the initial degree of mass segregation, the
ratio $R/R_t$ and the amount of impulsive mass loss. As noted
above, for example, for clusters like {\it S-Rg1}  with
small dynamical times, the amount of impulsive mass loss is negligible
and the dissolution time
of a mass-segregated cluster is similar to that of an unsegregated cluster. 
On the other hand, for very large amounts of impulsive
mass loss---larger than those in the systems we have studied 
---the dissolution time
must converge to a value of $\sim t_{tid}$.  It is therefore not
possible to derive a single general power-law to fit the scaling
of $T_{diss}$ with $R_g$.

For the specific set of standard segregated {\it S} runs, and
limiting the fit to systems with $R_g=4,~18,~40$, the dissolution
time is much shorter than that of the unsegregated systems 
and has a weaker dependence on $R_g$, $T_{diss}\sim
R_g^{0.43}$. 

The dissolution times of the mass-segregated clusters undergoing rapid
dissolution do not depend on $N$, as shown by the results of
simulations starting with the same degree of segregation but larger
values of $N$ (see Fig.~\ref{fig:tdissol}).

\begin{figure}[h]
\begin{center}
 \plotone{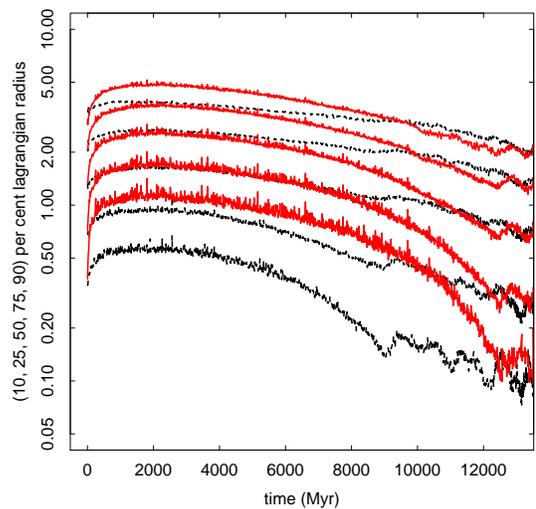} 
 \caption{Time evolution of the 10, 25, 50, 75, 90 percent Lagrangian
 radii for {\it S-Rg18R05}  (red solid line) and {\it
 U-Rg18R05} (black dashed line) at $R_g=18$ kpc.}
   \label{fig:lagrRg18R05}
\end{center}
\end{figure}

The difference in response of segregated and unsegregated clusters to
the same amount of impulsive mass loss is further illustrated by
Fig.~\ref{fig:lagrRg18},
which shows the time evolution of the 10, 25, 50, 75, 90 percent
Lagrangian radii for  {\it S-Rg18} and {\it U-Rg18}. 
Both clusters initially lose the same fraction 
of their initial mass and expand in response to  
this mass loss; however the preferential removal of mass from the
innermost regions in the segregated  cluster, leads to a much stronger
expansion and to rapid cluster dissolution. 

The longer dissolution times {\it L-Rg18} and  {\it M-Rg18}  plotted in
Fig.~\ref{fig:tdissol} show that, as the initial degree of mass
segregation decreases, so does the disruptive effect of the early mass
loss due to SN ejecta. 

The long-lived globular clusters we observe today might have
had very low degrees of initial mass
segregation which allowed them to survive.
Alternatively, a cluster with a higher
degree of initial mass segregation can survive if its initial size is
significantly smaller than its tidal radius.  In this case,
most cluster stars remain 
within the tidal boundary during the early expansion,
and the number of stars lost during this phase is smaller.

To illustrate this point, we have repeated the  {\it
S-Rg18} and {\it U-Rg18} runs,  but setting the ratio of the initial
cluster size, $R$, to the tidal radius $R_t$ equal to 0.75 ({\it
  S-Rg18R075} and {\it U-Rg18R075})  and 0.5 ({\it
  S-Rg18R05} and {\it U-Rg18R05}) . 
The dissolution times for these  systems are plotted in
Fig.~\ref{fig:tdissol} and show that the lifetime of a mass-segregated cluster
with the same degree of initial segregation increases as the ratio
$R/R_t$ decreases. 

In particular, the lifetimes of {\it S-Rg18R05} and  {\it
  U-Rg18R05} are very similar, as the early mass 
loss disruptive effects for the initially mass-segregated system are 
strongly suppressed by the fact that the cluster is expanding well
within its tidal 
radius and not immediately losing stars as it expands.
Since the system is initially denser and has a shorter dynamical time,
the amount of impulsive mass loss is smaller than for the tidally
truncated system; this effect also contributes, 
although to a lesser extent, to the extension of the cluster lifetime.

The evolution of the 10, 25, 50, 75, 90
percent Lagrangian radii for  {\it S-Rg18R05} and  {\it
  U-Rg18R05} are shown in Fig.~\ref{fig:lagrRg18R05}.
Although the lifetimes of these two systems are similar,
Fig.~\ref{fig:lagrRg18R05} clearly shows important differences in
the structural evolution of the two clusters: the mass-segregated cluster
undergoes a stronger initial expansion which, although it does not
lead to its prompt dissolution, significantly delays the
cluster core collapse and keeps the cluster
concentration lower than  that of the unsegregated cluster. 
\begin{figure}[h]
\begin{center}
 \plotone{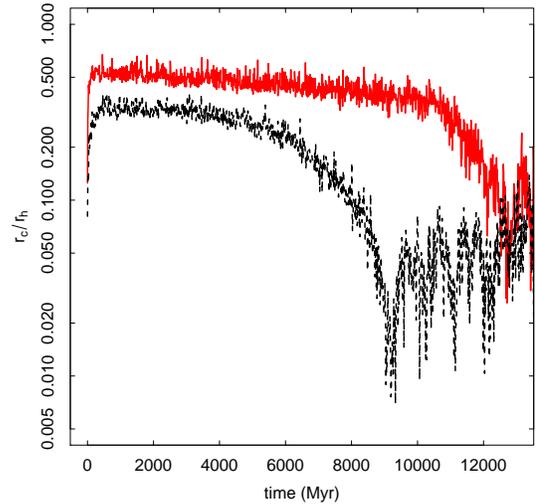} 
 \caption{Time evolution of the ratio of the core to the half-mass radius, $r_c/r_h$, 
for  {\it S-Rg18R05} (red solid line), {\it U-Rg18R05}
 (black dashed line). }
   \label{fig:conc}
\end{center}
\end{figure}
This is further
illustrated by Fig.~\ref{fig:conc} which shows the time   
evolution of the cluster concentration as measured by the ratio of
the core and half-mass radii, $r_c/r_h$, for 
 {\it S-Rg18R05} and {\it U-Rg18R05}.

\subsection{Results: evolution of the stellar mass function}
\label{sec:imf}
An important issue related to a cluster mass loss is the evolution
of the stellar MF. It has been  
shown in a number of investigations of initially unsegregated clusters
(see e.g. Vesperini \& Heggie 1997, Baumgardt \& Makino 2003) that mass loss due to two-body
relaxation leads to the preferential escape of low-mass stars and to the
flattening of the stellar MF. 

In particular, for long-lived clusters  losing mass  due to two-body
relaxation the extent of the flattening of the slope of  
the stellar MF is strictly correlated to the fraction
of the initial cluster mass left (see e.g.~Fig.~16 in Vesperini \&
Heggie 1997 and Eq.~14 in Baumgardt \& Makino 2003).

For long-lived initially unsegregated clusters, the dynamical mass loss
timescale is closely related to the two-body relaxation timescale and is, in
general, much longer than the stellar evolution timescale of the most
massive stars.  In this case, the range of stellar masses involved in
segregation and evaporation 
processes after a few billion years
includes the most massive dark remnants (1.2-1.4 $m_{\odot}$),
the more numerous white dwarfs, and the remaining main
sequence stars ($m \ltorder 1m_{\odot}$).
The most massive main sequence stars usually included in the
calculation of the slope of the mass function 
fall in the high-mass end of this range, and hence tend to sink to  the
cluster center, while low-mass main sequence
stars (e.g. $m\sim 0.1-0.2~m_{\odot}$) preferentially
escape from the cluster, flattening the MF.

For an initially mass-segregated cluster, on the other hand, the range of
stellar masses affected by segregation is 
much broader.  The most
massive stars in this broad range preferentially populate the
innermost regions, while the long-lived main-sequence stars
with ($m \ltorder 1  m_{\odot}$) all have very similar spatial distributions.

\begin{figure}[h]
\begin{center}
 \plotone{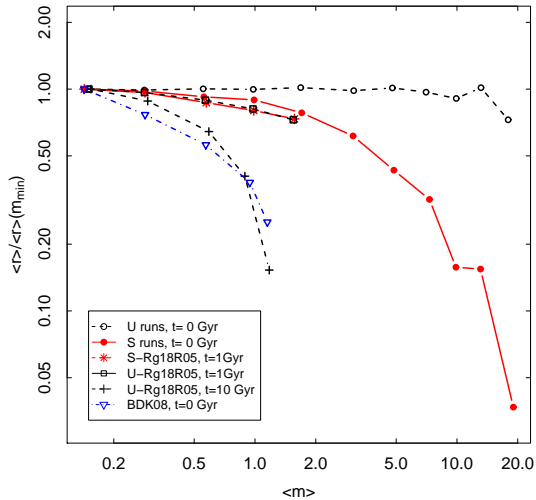} 
 \caption{Mean radial distance from the center of the cluster of stars
 (normalized to the mean radial distance of stars  
 in the first bin) binned in mass versus the mean  mass of
 stars in each bin. The different lines show (from the top to the bottom line) 
the {\it U} runs at $t=0$ (dashed line-open dots), the {\it  S} runs
 (solid line-filled dots) at  $t=0$, 
 {\it  S-Rg18R05} at $t=1$ Gyr (red solid line-asterisk dots) {\it
 U-Rg18R05} at $t=1$ Gyr (black dashed line-open square dots) and {\it
 U-Rg18R05} (black dashed line-crosses) at $t=10$ Gyr (the lines for
 the models at $t=1$ Gyr are almost completely overlapped).  
The blue dotted  line-/open triangles shows the {\it initial}
 segregation of one of the models with initial segregation (model No.3
 in their Table 2) studied by Baumgardt, De Marchi \& Kroupa
 (2008) (BDK08). } 
   \label{fig:radmass}
\end{center}
\end{figure}

This point is illustrated in Fig. \ref{fig:radmass}, which
shows the initial mean distance of stars from the cluster center
as a function of the mean stellar
mass in a number of  mass bins
for the {\it S} and  {\it U} runs.
The same information later in the evolution of the {\it
  U-Rg18R05}, when a  
significant degree of mass segregation has been reached and the
cluster hosts stars with a narrower range of masses, is also shown.

Fig. \ref{fig:radmass} shows that the initial degree of segregation for
stars with masses in the range 
$0.1<m/m_{\odot}<0.5$ (hereafter we focus on the
evolution of the MF of main sequence stars with initial masses in this
range) is negligible and similar in clusters with and without
primordial segregation.  As discussed above, primordial mass
segregation affects mainly the most massive stars initially present in
the cluster.
The segregation profile attained later in the evolution by an
unsegregated cluster shown in Fig.~\ref{fig:radmass} clearly
illustrates the differences with the  
initial segregation profile of a cluster with primordial segregation.
This is an important point in understanding the implications
of  primordial mass segregation for the evolution of the MF.

As the results presented in the previous sections show, the fate of
initially mass-segregated clusters and the processes driving their
dissolution depend on the degree of initial mass segregation and on
a cluster's initial size relative to its tidal radius. 
Hereafter we focus on the {\it S-Rg18}, {\it S-Rg18R05},
{\it U-Rg18}, {\it U-Rg18R05} runs, to explore the difference
between the MF evolution in systems with and 
without primordial mass segregation. 

\begin{figure}[h]
\begin{center}
\epsscale{0.95}
 \plotone{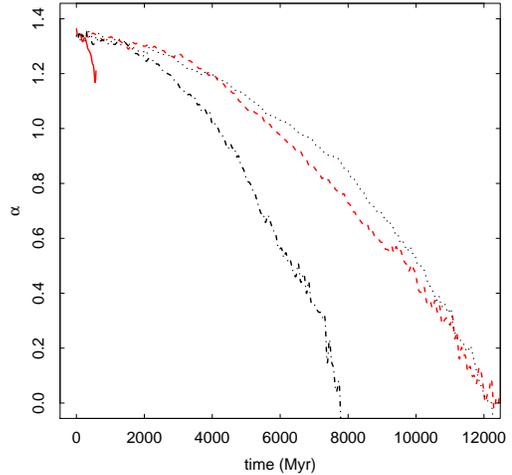} \\
 \plotone{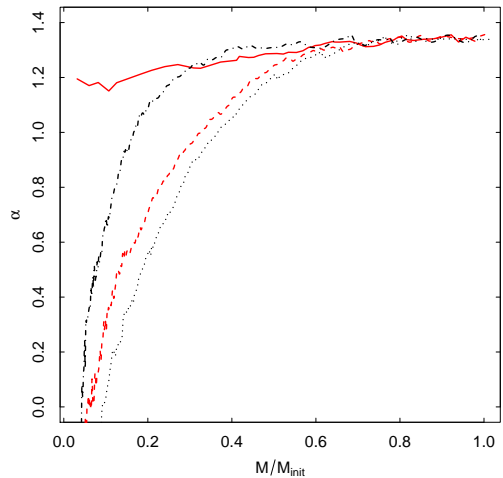} 
 \caption{({\it Upper panel} ) Time evolution of the slope of the mass function, $\alpha$,
   for main sequence stars with $0.1<m/m_{\odot}<0.5$  for {\it
   S-Rg18} (red solid line), {\it U-Rg18} (black dot-dashed line), {\it
   S-Rg18R05} (red dashed line) and {\it  U-Rg18R05} (black dotted line). 
   ({\it Lower panel}) $\alpha$ versus the fraction of the cluster
   initial mass left in a cluster (different lines refer to the same
   runs as in the upper panel).}  
   \label{fig:alpha}
\end{center}
\epsscale{1}
\end{figure}

We fit the mass function of main sequence stars with masses in
the range ($0.1<m/m_{\odot}<0.5$) with a power law 
$\hbox{d}N(m)=Am^{-\alpha} \hbox{d}m$;
the upper panel of Fig.~\ref{fig:alpha} shows the time evolution of
$\alpha$ for  {\it S-Rg18} and  {\it U-Rg18}.
 As discussed in the previous section, the
initially  mass-segregated system rapidly dissolves as its outer layers
expand in response to mass loss due to SN ejecta. As the system
quickly loses mass, the MF slope also rapidly flattens; the
evolution, mass loss, and MF flattening occur on a much longer
timescale for the unsegregated system. 

The lower panel in Fig.\ref{fig:alpha} 
shows the MF slope versus the fraction of the initial
mass left in the cluster, clearly illustrating an important
difference between mass-segregated and unsegregated clusters: for a given
amount of mass loss, the degree of MF flattening is significantly larger for the
initially unsegregated cluster.  This is due to the
differences in the range spanned by stellar masses during the cluster
evolution.  For the slowly evolving initially unsegregated cluster, the
narrower range of star masses present in the cluster during much of its
evolution results in greater preferential loss of
stars at the low-mass end of the range $0.1<m/m_{\odot}<0.5$,
while stars at the high-mass end of this range sink to the cluster center.
This is not the case for the initially mass-segregated system, in
which rapid dissolution ensures that stars much more massive than
$0.5 m_\odot$ are present during most of the evolution, preventing the
development of strong differences in the mass loss rate and the spatial
segregation of stars with masses in the range $0.1<m/m_{\odot}<0.5$. 

The time evolution of $\alpha$ and the  dependence of
$\alpha$ on the fraction of mass remaining in the cluster for the long-lived
{\it S-Rg18R05} and {\it U-Rg18R05} systems are also shown in
Fig.~\ref{fig:alpha}. The evolution of $\alpha$
is very similar for these two systems. 
The initial segregation profiles for  {\it U-Rg18R05} and {\it
  S-Rg18R05}  in Fig. \ref{fig:radmass} clearly show
that the degree of mass segregation for stars with masses in the range
$0.1<m/m_{\odot}<0.5$ is very similar in these two systems. If a
mass-segregated cluster survives its early expansion, as in the case of
{\it S-Rg18R05}, it will enter the evolutionary phase 
driven by two-body relaxation (after the most massive stars have
evolved off the main sequence) with a segregation profile similar to that of an
initially unsegregated cluster (see the mass segregation profiles of 
{\it   S-Rg18R05} and {\it U-Rg18R05}  at $t=1$ Gyr
in Fig.\ref{fig:radmass}), and the MFs of both systems
will evolve similarly.

For clusters initially underfilling
their Roche lobes, our simulations show that a given amount of mass
loss leads to a stronger MF  
flattening than for tidally truncated unsegregated clusters. 
This is a consequence of the slower mass loss rate of clusters initially
lying inside their tidal radii; it takes longer for these
clusters to lose a given amount of mass, and during this additional time
a higher degree of mass segregation for stars with masses
$0.1<m/m_{\odot}<0.5$ is reached.

\section{Discussion and Conclusions}
\label{sec:disc_and_concl}

The results of our study show that initial mass segregation can
significantly affect the dynamical evolution of star clusters.

For clusters initially filling their Roche lobes, our simulations
show that mass-segregated clusters can quickly dissolve as a result of
the early expansion triggered by  mass loss due to stellar
evolution, unless the initial degree of  mass
segregation is low. We have studied the evolution of  mass-segregated and
unsegregated clusters undergoing the same amount of early impulsive
mass loss and find that, for segregated clusters, 
the preferential removal of this mass from the innermost regions 
strongly augments the strength of the early
expansion and leads to rapid cluster dissolution.

We have also explored the evolution of clusters initially underfilling
their Roche lobes, showing that in this case mass-segregated clusters
can survive the early expansion phase; a mass-segregated cluster still
undergoes a strong expansion in 
this case, but much of this expansion occurs within its tidal
radius without significant loss of stars. The subsequent cluster evolution,
lifetime, and mass loss rate do not differ significantly from those of
an initially unsegregated cluster. 

The stronger expansion of the segregated cluster, while not leading to
rapid dissolution, does drive the cluster toward a less concentrated
structure, significantly delaying core collapse.  The less
concentrated structure of an initially segregated cluster throughout
most of the evolution is the main difference between long-lived
segregated and unsegregated clusters.  If clusters still surviving
today were initially segregated, this is one of the possible
evolutionary routes they might have followed.  Another possible
scenario has been studied by D'Ercole et al. (2008), who demonstrate
the role of primordial segregation in the evolution of clusters which
form a second generation of stars in the innermost regions of a
first-generation population.  The early expansion in this case plays a
key role in causing the loss of a large fraction of the
first-generation stars, while most of the centrally concentrated
second generation remains in the cluster, resulting in a system
containing a mix of first and second generation stars that eventually
survives until the present day.

We have also studied the differences in the evolution of the stellar
MF for segregated and unsegregated clusters, focusing on the evolution
of the MF of low-mass ($0.1<m/m_{\odot}<0.5$) main sequence stars.
Our simulations show that the MF of a rapidly dissolving segregated
cluster flattens 
quickly as the system loses stars. However the extent of the MF
flattening induced by a given amount of mass loss for a tidally
truncated segregated cluster (our {\it S} runs) is smaller
than for an initially unsegregated system (our {\it U} runs). 

The difference is a consequence of the rapid dissolution of segregated
clusters.  We have shown that the initial degree of segregation
adopted in our simulations
mostly affects massive stars with $m \gtorder 5-10 m_{\odot}$, while low-mass
stars with $m \ltorder 1 m_{\odot}$, initially have a similar
spatial distribution to stars in unsegregated clusters. 
Segregated clusters initially filling their Roche lobe  quickly lose
mass before two-body relaxation can
produce a significant difference in the spatial distribution (and in the
mass loss rate) of stars with masses $m \ltorder 1 m_{\odot}$.
Unsegregated clusters, on the other hand, lose stars more slowly and a
large fraction of their initial mass is lost late in the evolution,
when the most massive stars in the range $0.1<m/m_{\odot}<0.5$ are
significantly more concentrated than stars of lower masses.  
The differences between the {\it S} and {\it U} systems
in the degree of MF flattening for a given amount of mass loss
result from differences in the processes 
driving the cluster dissolution and, more importantly, from
differences in the dissolution timescales associated with these processes. 

Segregated clusters initially well confined within
their Roche lobe may survive their early expansion and, in this case,
we have shown that both segregated
and unsegregated clusters display very similar MF evolution.
As pointed out above, the main fingerprint of initial segregation might  be
looser cluster structure
during a larger fraction of the cluster lifetime.  However, we emphasize
that, unless some other mechanism continues to drive the cluster
expansion,
long-lived segregated clusters also eventually undergo core
collapse; therefore it can not be ruled out that high-concentration
and post-core collapse clusters might have been initially segregated \footnote{ See e.g. Mackey et al. (2007, 2008) for a study of the early expansion
of segregated clusters driven by stellar evolution mass loss and the
subsequent heating from a population of stellar black holes, and the
role these processes might play in determining the radius-age trend
observed for massive clusters in the Magellanic Clouds.}.

It has recently been suggested by Baumgardt \etal\ (2008) that the
low-concentration clusters with strongly flattened MFs found in the
observational study by De Marchi \etal\ (2007) can result only from
the dynamical evolution of initially segregated clusters (see also
Marks \etal\ 2008 for the possible role of primordial gas expulsion).
In particular, the simulations of Baumgardt et al. show that a cluster
must initially be significantly mass segregated {\em and} fill its
Roche lobe in order to evolve into a system with properties resembling
those of the loose, flat-MF clusters observed by De Marchi et al.  The
initial degree of segregation adopted by Baumgardt et al. is
significantly larger than that of our simulations, as shown in
Fig.\ref{fig:radmass}.  However, the Baumgardt et al. study focuses
on the effects of two-body relaxation, and does not include the early
mass loss due to stellar evolution which, as shown by our simulations,
can play a crucial role in cluster evolution.  As discussed above,
initially tidally limited segregated clusters may dissolve rapidly due
to the expansion triggered by early mass loss, even when starting with
a degree of initial segregation much smaller than that adopted by
Baumgardt et al.

In a future investigation we will focus on detailed comparisons
between our simulations and observational data, with the goal of
assessing which, if any, observed cluster properties might represent
unique signatures of initial mass segregation.

\acknowledgments
EV and SM were supported in part by NASA grants NNX07AG95G and
NNX08AH15G, and by NSF grant AST-0708299.  SPZ was supported by the
National Organization for Scientific Research (NWO, grant
\#643.200.503), the Netherlands Advanced School for Astronomy (NOVA)
and the Leids Kerkhoven Bosscha Fonds (LKBF).  We thank H. Baumgardt
for sending us the data of one of the initial conditions used in
Baumgardt et al. (2008).

\end{document}